\documentclass[conference]{IEEEtran}
\makeatletter
\def\ps@headings{%
\def\@oddhead{\mbox{}\scriptsize\rightmark \hfil \thepage}%
\def\@evenhead{\scriptsize\thepage \hfil \leftmark\mbox{}}%
\def\@oddfoot{}%
\def\@evenfoot{}}
\makeatother
\usepackage[utf8]{inputenc}
\usepackage{amsmath}
\usepackage{booktabs}
\usepackage{listings}
\usepackage{cite}
\usepackage{adjustbox}
\usepackage{xcolor}
\usepackage{color,soul}
\usepackage{multirow}
\usepackage{enumerate}
\usepackage[
  separate-uncertainty = true,
  multi-part-units = repeat
]{siunitx}
\usepackage{algorithm}
\usepackage{graphicx}
\usepackage{textcomp}
\usepackage{caption}
\usepackage{subcaption}
\usepackage{authblk}
\usepackage{caption}
\usepackage{amsfonts}
\usepackage{algpseudocode}
\usepackage{url}
\usepackage[justification=centering]{caption}

\definecolor{colorEntityBack}{rgb}{0.01, 0.01, 0.4}
\definecolor{colorPolicyBackDarkDark}{rgb}{0.81, 0.81, 0.86}
\definecolor{ashgrey}{rgb}{0.7, 0.75, 0.71}
\definecolor{beaublue}{rgb}{0.74, 0.83, 0.9}
\definecolor{hotpink}{rgb}{1.0, 0.41, 0.71}
\definecolor{darkgreen}{rgb}{0.09, 0.45, 0.27}
\definecolor{cerulean}{rgb}{0.105, 0.672, 0.836}
\definecolor{candyapplered}{rgb}{1.0, 0.03, 0.0}
\definecolor{cadetgrey}{rgb}{0.57, 0.64, 0.69}
\definecolor{trolleygrey}{rgb}{0.5, 0.5, 0.5}

\begin{document}

\date{18 February 2024}

\title{CND-IDS: Continual Novelty Detection for Intrusion Detection Systems}

\author[1]{Sean Fuhrman}
\author[2]{Onat Gungor}
\author[2]{Tajana Rosing}

\affil[1]{Department of Electrical and Computer Engineering, University of California, San Diego}
\affil[2]{Department of Computer Science and Engineering, University of California, San Diego}
\affil[ ]{\textit{\{stfuhrma, ogungor, tajana\}@ucsd.edu}}

\maketitle
\pagestyle{plain}
\pagenumbering{gobble}
\newcommand{\norm}[1]{\left\lVert#1\right\rVert}
\newcommand{\Design}[0]{CND-IDS}

\begin{abstract}
Intrusion detection systems (IDS) play a crucial role in IoT and network security by monitoring system data and alerting to suspicious activities. Machine learning (ML) has emerged as a promising solution for IDS, offering highly accurate intrusion detection. However, ML-IDS solutions often overlook two critical aspects needed to build reliable systems: continually changing data streams and a lack of attack labels. Streaming network traffic and associated cyber attacks are continually changing, which can degrade the performance of deployed  ML models. Labeling attack data, such as zero-day attacks, in real-world intrusion scenarios may not be feasible, making the use of ML solutions that do not rely on attack labels necessary. To address both these challenges, we propose \Design{}, a continual novelty detection IDS framework which consists of (i) a  learning-based feature extractor that continuously updates new feature representations of the system data, and (ii) a novelty detector that identifies new cyber attacks by leveraging principal component analysis (PCA) reconstruction. Our results on realistic intrusion datasets show that \Design{} achieves up to 6.1$\times$ F-score improvement, and up to 6.5$\times$ improved forward transfer over the SOTA unsupervised continual learning algorithm. Our code will be released upon acceptance. 

\end{abstract}


\section{Introduction}

In today’s digital landscape, cybersecurity is essential for safeguarding sensitive data and maintaining trust in digital systems. With cyber threats becoming more sophisticated, organizations must adopt thorough security measures to protect against breaches and unauthorized access \cite{emerging2023kumar}. A critical part of such security measures are intrusion detection systems (IDS) which identify malicious network traffic and computer usage \cite{liao2013intrusion}. Recently, Machine Learning (ML) has become a popular IDS solution due to its great attack detection performance and less requirement for human knowledge \cite{gungor2024roldef}. However, state-of-the-art (SOTA) ML-IDS solutions often fail to adequately address the challenges posed by (i) the evolving nature of cyber attacks and (ii) the lack of attack labels.

A University of Maryland survey predicts that a new cyber attack happens somewhere on the Internet every 39 seconds \cite{hackerstudy2007}, underscoring the need for dynamic IDS. 
However, most SOTA ML-IDS approaches are static (offline) 
and cannot easily adapt their behavior over time \cite{verwimp2023continual}. ML models, when taught new knowledge, \textit{catastrophically forget} previously known information. To efficiently address this problem, continual learning (CL) is proposed which is capable of learning from an infinite stream of data \cite{wang2024comprehensive}. 
An experience is defined as a shift in the data stream distribution, and CL models are designed to function well across multiple experiences. This capability is currently essential in cybersecurity, where each new attack type can be considered as a new experience\cite{kozik2019balanced}. To illustrate, an intrusion detection model trained for a group of attacks (experiences) such as \textit{worm}, \textit{shellcode}, and \textit{exploits} should be able to identify new types of attacks (experiences) such as \textit{backdoor} and \textit{fuzzers} \cite{karn2021learning}. Some studies have adopted CL for intrusion detection \cite{kozik2019balanced, karn2021learning, prasath2022analysis, kumar2023augmented, oikonomou2023multi, channappayya2024augmented}, but they rely on labeled attack data, which may not always be feasible to obtain. This causes the ML model to overfit to common attacks in the dataset, resulting in a failure to detect rare and zero-day threats.
\begin{figure}
    \centering
    \includegraphics[width=\linewidth]{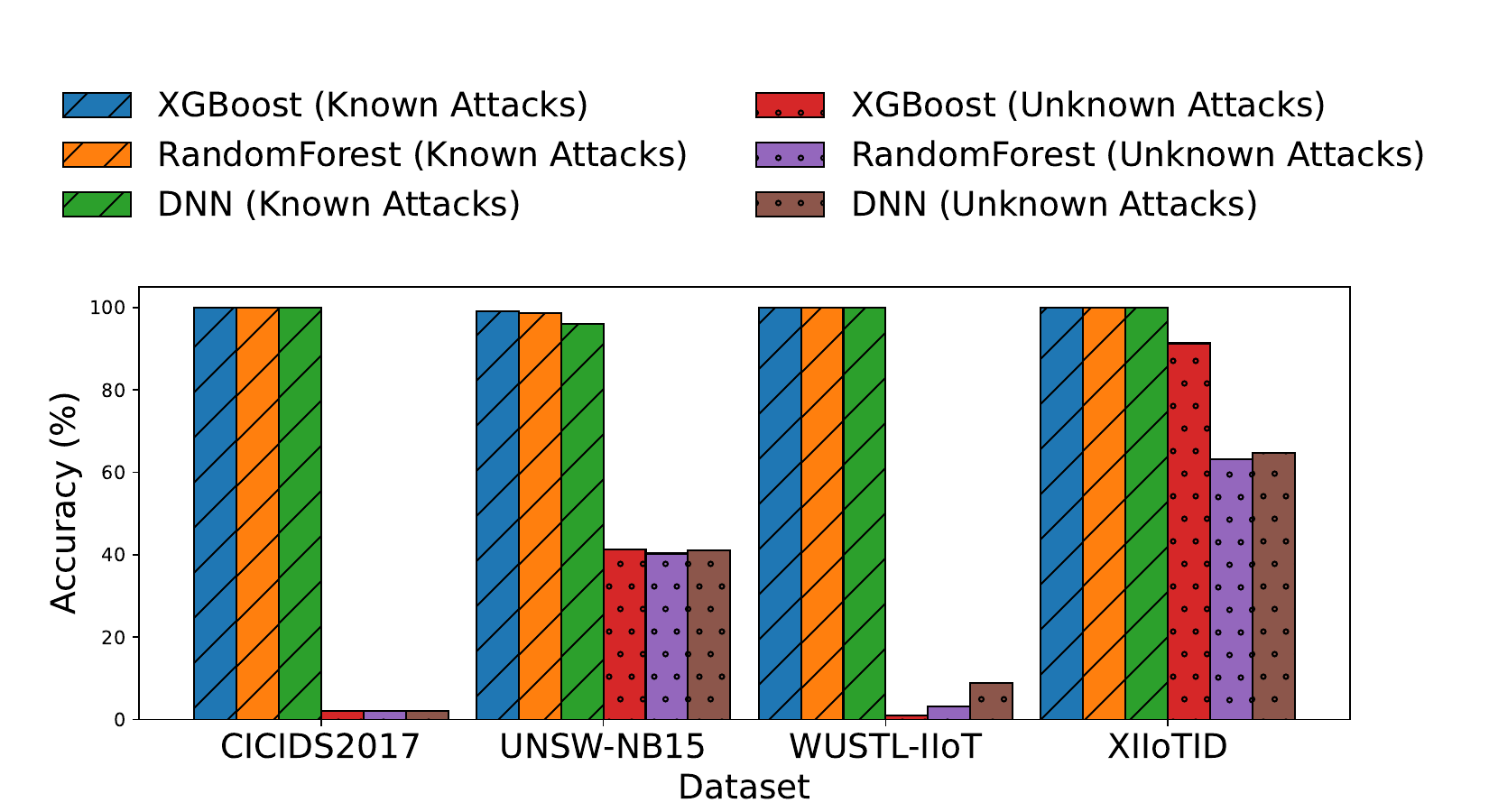}
    \caption{State of the art machine learning intrusion detection performance on known/unknown attacks}
    \label{fig:unknown_attacks}
\end{figure}

\begin{figure*}
    \centering
    \includegraphics[width=\linewidth]{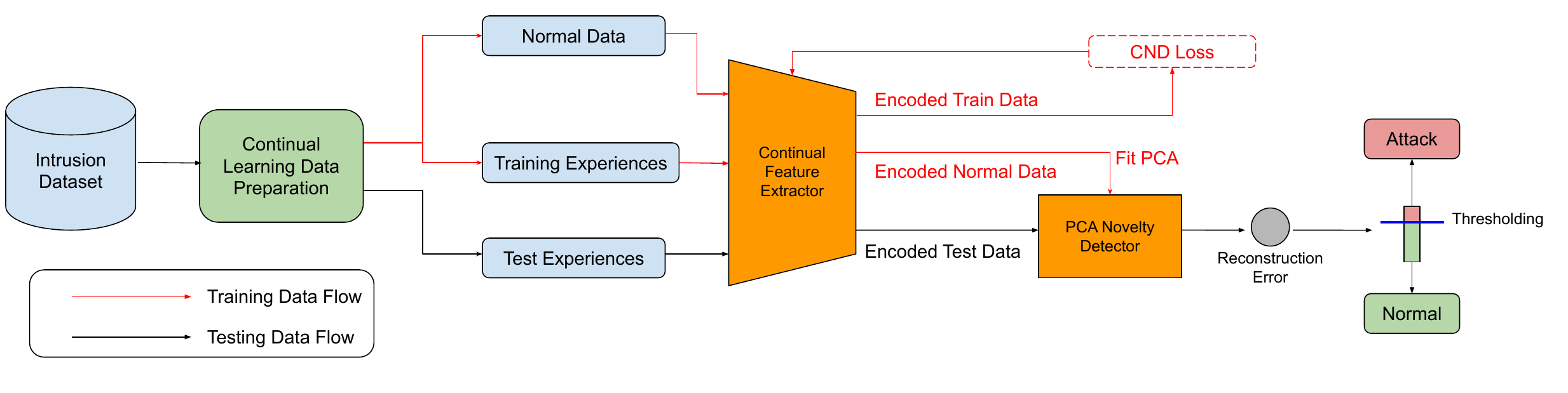}
    \caption{Continual Novelty Detection IDS (\Design{}) Framework}
    \label{figure:model_diagram}
\end{figure*} 


SOTA ML-IDS solutions have been shown to have high accuracy on well represented attacks within the data, however have low precision on infrequent attacks \cite{macas2022survey}. It is extremely difficult for zero-day attacks to be well represented in the data. Figure~\ref{fig:unknown_attacks} illustrates the limitations of current ML-IDS methods in handling zero-day attacks, revealing a substantial decline in performance when faced with previously unseen threats. To address this issue, novelty detection (ND)-based ML methods have demonstrated significant promise in detecting both known and zero-day attacks without requiring a labeled dataset \cite{verkerken2022towards}. 
These ND solutions can effectively identify attacks by analyzing purely normal data, provided that the attacks exhibit substantial deviations from normal behavior \cite{verkerken2022towards}. 
However, it has been shown that ND algorithms such as Isolation Forest and Local Outlier Factor
result in lower prediction performance under changing environments, where they suffer from issues like \textit{catastrophic forgetting} \cite{Faber_2024}. 
This issue can be addressed by leveraging continual learning strategies, hence motivating continual novelty detection methods for IDS. 

To address the challenges of evolving data streams and lack of supervision, we propose \textit{\Design{}}, a continual novelty detection approach for ML-based intrusion detection (Fig.~\ref{figure:model_diagram}). To the best of our knowledge, we are the first to propose continual ND for intrusion detection. 
\Design{} consists of two critical components: a continual feature extractor (CFE) and a novelty detector based on principal component analysis (PCA) reconstruction. We design a novel continual ND loss to learn representations of the data stream, enabling the CFE to update its representations in a streaming environment with zero attack labels. 
PCA reconstruction error is then used to classify test samples as either attack or normal data based on thresholding. By fitting PCA on normal data, \Design{} is trained without any labeled attack data, enabling it to detect previously unknown zero-day attacks. Our results show that on realistic intrusion datasets, \Design{} achieves up to 6.1$\times$ F-score improvement over the SOTA unsupervised continual learning algorithm.

\section{Related Work}
\textbf{Intrusion Detection Systems:} 
Intrusion Detection Systems (IDS) fall into two main categories: Signature-based and Anomaly-based IDS. Signature-based IDS identify attacks by matching network data to known attack patterns. In contrast, Anomaly-based IDS use machine learning methods to analyze statistical properties of network data \cite{gupta2023surveyIDS}. Machine learning-based IDS (ML-IDS), a popular form of Anomaly-based IDS, use historical data to train models, enhancing accuracy and reducing the need for expert knowledge \cite{gungor2024rigorous}. The literature proposes various ML models, including supervised models such as support vector machine, random forest, deep neural networks \cite{liu2019machine}, and unsupervised models like isolation forest, autoencoder, and principal component analysis \cite{verkerken2022towards}.

\textbf{Continual Learning for Intrusion Detection:} Current ML-IDS solutions are often built on the premise that all data are sampled \textit{i.i.d.} (independent and identically distributed) \cite{verwimp2023continual}, making them vulnerable to performance degradation when data distributions shift. 
Continual Learning (CL) seeks to train ML models to acquire new information while minimizing the loss of previously learned knowledge. 
Prasath et al. \cite{prasath2022analysis} highlight that intrusion datasets exhibit these shifting distributions, underscoring the need for CL-capable IDS that can autonomously adapt to the evolving characteristics of the data.
Few studies have recently adopted CL for intrusion detection. Kozik et al. \cite{kozik2019balanced} extend efficient lifelong learning algorithm \cite{ruvolo2013ella} to detect cyber attacks and address the data imbalance problem. 
Kumar et al.\cite{amalapuram2022continual} applied Elastic Weight Consolidation and Gradient Episodic Memory CL algorithms. In a follow-up study \cite{kumar2023augmented}, they extend a memory-based CL method and proposed a novel approach (perturbation assistance for parameter approximation), improving CL scalability. Nevertheless, all of these studies leverage labeled attack data to devise their CL solutions, which may not be feasible in IDS due to the difficulty and expense of obtaining such labeled data.     

\textbf{Continual Novelty Detection:} 
Continual Novelty Detection (CND) primarily addresses detecting distribution shifts after ML model deployment. To illustrate, Kim et al. \cite{kim2023openworld} highlight how ND combined with CL is necessary for realistic class incremental learning. 
Aljundi et al. \cite{aljundi2022continual} investigate CND for detecting data outside the current model's knowledge. Finally, Rios et al. \cite{rios2022incdfm} propose a novel CND framework for detecting new data and incorporating it into the current ML model's knowledge. 
These studies focus on using ND to identify data outside the training set and integrate it into the ML model via CL strategies. However, Faber et al.\cite{Faber_2024} demonstrate the performance decline in ND-algorithms themselves under continual learning settings. 
Given the constantly changing nature of attacks, it is necessary for ML-IDS systems to continually adapt without labels, which motivates our work. 
To the best of our knowledge, we are the first to propose continual novelty detection for ML-based intrusion detection.

\section{\Design{} Framework}
\label{framework}
Fig.~\ref{figure:model_diagram} illustrates our continual novelty detection-based intrusion detection (\Design{}) framework. We first apply a continual learning data preparation phase, where we divide the intrusion data into normal data, training experiences, and test experiences. Our framework consists of two main components: a continual feature extractor (CFE) and principal component analysis (PCA)-based novelty detector. The CFE creates a representation of the data that is used to distinguish normal and attack data by the novelty detector. The CFE allows the model to learn new feature representations overtime, so that it can easily adapt as new attacks are introduced through the data stream. CFE is trained through our continual novelty detection loss function, to continually learn from the input stream without forgetting. The next component, PCA reconstruction, provides an anomaly score on the feature output from the feature extractor. PCA is trained purely on normal data, therefore it can generalize well to unseen attacks. Finally we apply a threshold, $\tau$, such that if the anomaly score is greater than $\tau$, we classify that data as an attack. Overall, our learning framework does not require any labeled attack data.

\subsection{Continual Learning (CL) Data Preparation}
\label{section:data_preparation}


We assume a given dataset contains attack data $A$, normal (benign) data $N$, and various attack types $C$. 
To begin our data preparation, we remove 10\% of the normal data, and save this as clean normal data $N_c$ to train PCA novelty detector. For the remaining data, we split it across an arbitrary number of experiences $m$. We create a set of experiences $E$, such that each individual experience is $(E_0, E_1,...E_{m})$. Each experience contains a portion of the remaining normal data, of size $\frac{0.9 *|N|}{m}$. The attacks are distributed such that each experience contains $\frac{|C|}{m}$ different attack classes. Therefore, each experience contains some number of \textit{unique} attacks to that experience. This allows us to test on zero-day attacks through future experiences, and seen attacks through current or past experiences. Each experience $E_i$ is split into training and testing set. The training split contains only $X_{train}$ data points, meaning it does not include any labels. The test data contains $\{X_{test},Y_{test}\}$ where $Y_{test} \in \{0,1\}$, where 0 means normal, and 1 means attack data. For each experience $E_i$, we utilize the training set $X_{train}^i$ and $N_c$ to train \Design{}. We then evaluate the model using the test points $\{X_{test},Y_{test}\}$. This setup is designed to represent a realistic intrusion detection scenario, where attacks are entirely unknown before deployment into a continual learning environment. In our framework, the only known data is $N$, while all attack data is unlabeled. With this data setup, we can assess model's performance across various attack scenarios. Specifically, we can simulate the model's response to zero-day attacks as well as known attacks.

\subsection{Proposed Algorithm}

\begin{algorithm}
    \caption{\Design{} Algorithm (Training and Test)}
    \label{alg:train_test}
    \begin{algorithmic}[1]
    \Require $E$ - Experience Set; $N_c$ - Subset of Clean Normal Data; $CFE$ - Continual Feature Extractor; $ND$ - PCA Novelty Detector
    \For {$i \leftarrow 1$ \textbf{to} $|E|$}  
      \State Get $X_{\text{train}}$ from experience data $E_i$
      \State Fit $CFE$ to training data $X_{\text{train}}$
      \State Encode $N_c$ by passing it through $CFE$
      \State Fit $ND$ to encoded $N_c$
      \State Get $X_{\text{test}}, Y_{\text{test}}$ from all experiences within $E$
      \State Encode $X_{\text{test}}$ by passing it through $CFE$
      \State Use $ND$ to compute anomaly score $S_{\text{test}}$ on test data
      \State Compute threshold $\tau$ based on thresholding method
      \State Compute predictions such that $Y_{\text{pred}} = (S_{\text{test}} > \tau$)
      \State Compute evaluation metrics based on $Y_{\text{pred}}, Y_{\text{test}}$
    \EndFor
    \end{algorithmic}
\end{algorithm}
Algorithm~\ref{alg:train_test} depicts the proposed \Design{} algorithm, including the training and test steps. The training flow follows three main steps: (i) train the CFE, (ii) encode $N_c$ into features, and (iii) train PCA-based novelty detector on the encoded $N_c$ data. Each of these steps is performed at every training experience. For step (i), the CFE is trained on $X_{train}$ using the loss function from Equation~\ref{equation:final_loss}. For step (ii), the subset of normal data $N_c$ is encoded by passing it through the trained CFE, where we encode the entire set. Ultimately, in step (iii), we fit the PCA novelty detection model to this encoded normal data. This completes the \Design{} training, and we then proceed to evaluate on the test experiences. At the end of each training experience, we evaluate on the test set of all experiences. Therefore, we evaluate model performance on unseen attacks (future experiences) and seen attacks (past experiences). When evaluating a batch of testing data, we encode all points from $X_{test}$ with CFE. Then use PCA reconstruction to get an anomaly score $S_{\text{test}}$ for each point based on the reconstruction loss. Next, we leverage widely used Best-F \cite{su2019robust} to select a threshold $\tau$. Ultimately, all anomaly scores greater than $\tau$ are classified as attack while all scores less than or equal to $\tau$ are classified as normal, resulting in the CND-IDS predictions $Y_{\text{pred}}$. We ultimately compute our evaluation metrics by comparing $Y_{\text{pred}}$ with the true labels $Y_{\text{test}}$.   

\subsection{Continual Feature Extractor (CFE)}
CFE is an autoencoder (AE)-based model that leverages Multi-Layer Perceptron (MLP) as both the encoder and decoder. 
Our continual novelty detection loss ($L_{CND}$) consists of a novel cluster separation loss ($L_{CS}$), a reconstruction loss ($L_{R}$), and a continual learning loss ($L_{CL}$): 

\begin{equation}
    \label{equation:final_loss}
    L_{CND} = L_{CS} + \lambda_R L_R + \lambda_{CL} L_{CL}
\end{equation}

where $\lambda_R \in [0,1]$ and $\lambda_{CL} \in [0,1]$ are hyper-parameters controlling the strength of the reconstruction loss and the continual learning loss respectively. 

\textbf{Cluster Separation Loss:} We design a novel cluster separation loss $L_{CS}$ to enhance the performance of our PCA-based novelty detector by increasing the separation between normal and anomalous data points in feature space. To achieve this, we leverage the clean normal data $N_c$ (which is also used to train the PCA model). Using K-Means clustering \cite{kmeans2022}, we identify which points in $X_{train}$ are most similar to $N_c$ and then push these points apart in feature space. 
Specifically, $L_{CS}$ leverages K-Means clustering to assign pseudo-labels to all points in $X_{train}$. Then it utilizes triplet margin loss\cite{schroff2015facenet} to maximize the Euclidean distance between the different pseudo labels. Calculating the $L_{CS}$ pseudo-labels involves the following steps: 1) Fit K-Means clustering to all points in $X_{train}$; 2) Find the cluster labels of all $N_c$ data points; 3) Create the set of "normal data" clusters $CL_N$, where each cluster contains at least one point from $N_c$; 4) Assign class 0 to points in $X_{train}$ if their associated cluster is in $CL_N$, and class 1 otherwise.
In summary, after fitting K-Means to $X_{train}$, we identify the clusters associated with $N_c$. If a cluster contains any point from $N_c$, all points in that cluster are assigned to class 0; otherwise, they are assigned to class 1. This effectively splits the data into two, where 0 would represent normal pseudo-label, and 1 would represent anomalous pseudo-label

After assigning the pseudo-labels, the final loss is computed using the widely adopted triplet margin loss \cite{schroff2015facenet}, defined as:
\begin{equation}
    L_{CS} = \max\left( \Delta_{ap} - \Delta_{an} + m, 0 \right)
\end{equation} where $\Delta_{ap}$ represents the Euclidean distance between an anchor point and a positive point (a point of the same class), and $\Delta_{an}$ is the Euclidean distance between the anchor and a negative point (a point from a different class). The term $m > 0$ is a hyper-parameter that specifies the desired margin between positive and negative points.

\textbf{Reconstruction Loss:} The reconstruction loss ensures that the features embedded by the encoder retain significant information from the original data. We found that this approach helps the PCA extractor better generalize to future and past experiences by forcing the model to learn encodings that align more closely with the original data. The reconstruction loss is the mean squared error (MSE) between the original input and the reconstructed output of the decoder. Let $h$ be the encoded feature of point $x$. Then $L_R = MSE(\text{decoder}(h) , x)$.

\textbf{Continual Learning Loss:} Our solution to catastrophic forgetting is employing a latent-based regularization loss, $L_{CL}$ \cite{ashfahani2023unsupervised}. This loss ensures that as the model learns new information, it retains knowledge from past experiences.  $L_{CL}$ can be formulated as: $
    L_{CL} = \sum_{i < c}^{c} MSE(h^{c}, h^{i}) $
where $c$ represents  the current experience, and \textbf{$h^{j}$}  is the encoded embedding of input $x$ at experience $j$. The loss is computed by summing the MSE between the current embedding (${h^c}$) and all previous embeddings $h^{i}$ where $i < c$. This ensures that the current embedding is still similar to the previous embedding, thereby preventing catastrophic forgetting. To calculate past embedding, we pass the current data point $x$ into a past version of the model. This requires the model to save its state between experiences but does not require it to save any data, which can significantly reduce storage overhead.
\subsection{PCA-based Novelty Detection}
Inspired by \cite{rios2022incdfm}, we create a PCA-based novelty detection approach. Let $\textbf{h}$ denote the output feature from our feature extractor. We utilize principal component analysis transformation $\textbf{T}$ to map the input feature from a high dimensional space to low dimensional space $\textbf{T} : \textbf{h} \rightarrow \textbf{l}$. We then utilize the feature reconstruction error ($FRE$) as the anomaly score: $FRE = \norm{\textbf{h} - \textbf{T}^{-1}(\textbf{l})} ^2$, where $\textbf{T}^{-1}$ is the inverse PCA transform. The anomaly score is therefore the reconstruction loss from this method. We train the PCA transformation on the subset of normal data $N_c$ after it is encoded by the CFE.  

\section{Experimental Analysis}
\label{experimental}
\subsection{Experimental Setup}
\label{section:experimental_setup}
\textbf{Datasets:} Table~\ref{tab:datasets} provides a detailed breakdown of the SOTA intrusion datasets utilized in our study. 
For IIoT intrusion, we use IIoT datasets X-IIoTID \cite{al2021x} and WUSTL-IIoT \cite{zolanvari2021wustl}. We also include commonly used network intrusion datasets CICIDS2017 \cite{Sharafaldin2018TowardGA} and UNSW-NB15 \cite{moustafa2015unsw}. For X-IIoTID \cite{al2021x}, CICIDS2017 \cite{Sharafaldin2018TowardGA}, and UNSW-NB15 \cite{moustafa2015unsw}, we split the data across five experiences such that each experience contains two to four attacks. For WUSTL-IIoT \cite{zolanvari2021wustl}, we split the data across four experiences such that each experience contains one attack. We perform this data split to simulate an evolving data stream with emerging cyber attacks over time where each experience contains different attacks.

\begin{table}[h]
    \caption{Selected Intrusion Datasets}
    \centering
    \label{tab:datasets}
    \resizebox{.99\columnwidth}{!}{
    \begin{tabular}{c|c|c|c|c}
    \hline
    Dataset    & Size      & Normal Data & Attack Data & Attack Types \\ 
    \hline
    X-IIoTID \cite{al2021x}   & 820,502   & 421,417     & 399,417     & 18           \\
    \hline
    WUSTL-IIoT \cite{zolanvari2021wustl} & 1,194,464 & 1,107,448   & 87,016      & 4       \\
    \hline
    CICIDS2017 \cite{Sharafaldin2018TowardGA} & 2,830,743 & 2,273,097 & 557,646 & 15 \\
    \hline
    UNSW-NB15 \cite{moustafa2015unsw}
 & 257,673 & 164,673 & 93,000 & 10 \\
    \hline
    \end{tabular}}
\end{table}

\textbf{Baselines:} 
We evaluate our algorithm against two SOTA unsupervised continual learning (UCL) algorithms: the Autonomous Deep Clustering Network (\textbf{ADCN}) \cite{ashfahani2023unsupervised}, and an autoencoder paired with K-Means clustering. The autoencoder K-Means model is combined with Learning without Forgetting \cite{lwf2019Li} continual learning loss; we refer to this model as \textbf{LwF}. Note that both \textbf{ADCN} and \textbf{LwF} require a small amount of labeled normal and attack data to perform classification. We also compare our approach against SOTA ND methods: local outlier factor (\textbf{LOF})\cite{Faber_2024}, one-class support vector machine (\textbf{OC-SVM})\cite{Faber_2024}, principal component analysis (\textbf{PCA})\cite{rios2022incdfm}, and Deep Isolation Forest (\textbf{DIF}) \cite{xu2023deep}. 
Since these ND models cannot be retrained on unlabeled contaminated data, continual learning is not feasible for these methods.


\begin{figure*}
    \centering
    \includegraphics[width=.95\linewidth]{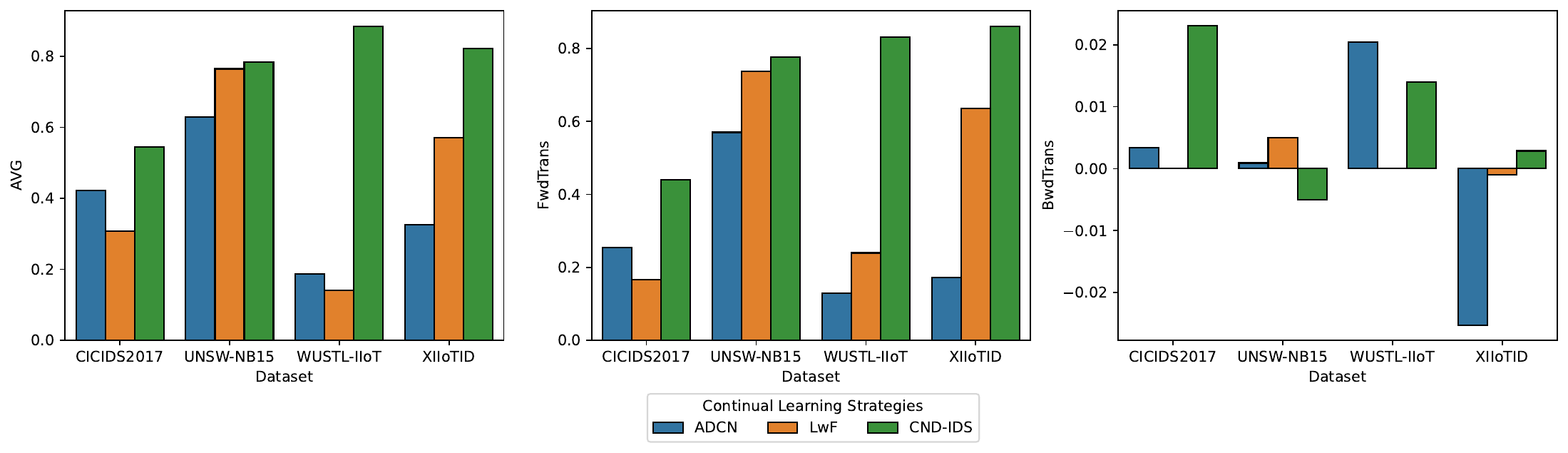}
    \caption{Continual learning metric results of ADCN\cite{ashfahani2023unsupervised}, LwF\cite{lwf2019Li}, and \Design{}}
    \label{fig:continual_methods_results}
\end{figure*}

\textbf{Evaluation Metrics:} To evaluate the model performance, we report $F_{1}$ score. Since there is a class imbalance within these datasets, to simulate real world IDS, $F_{1}$ score gives an accurate idea on attack detection. For the continual learning methods, we evaluate their performance at the end of each training experience on all experience test sets. This generates a matrix of $F_{1}$ score results $R_{ij}$ such that $i$ is the current training experience, and $j$ is the testing experience. To summarize this matrix of results, we report widely used CL metrics \cite{diaz2018don}: average $F_{1}$ score on current experience (AVG), forward transfer (FwdTrans), and backward transfer (BwdTrans). For a matrix $R_{ij}$ with $m$ total experiences, our metrics are formulated as follows: $\text{AVG}_{F_1} = \frac{\sum_{i = j} R_{ij}}{m}$; $\text{FwdTrans}_{F_1} = \frac{\sum_{j>i} R_{ij}}{\frac{m * (m-1)}{2}}$; $\text{BwdTrans}_{F_1} = \frac{\sum_{i}^m R_{mi} - R_{ii}}{\frac{m * (m-1)}{2}}$.
AVG is the average performance on the current test experience at every point of training. FwdTrans is the average performance on ``future'' experiences, which simulates performance on zero-day attacks. Finally, BwdTrans is the average change in performance of ``past'' test experiences at a ``future'' point of training. A negative BwdTrans indicates catastrophic forgetting, whereas a positive BwdTrans  indicates the model actually improved performance on past experiences after learning a future experience. Overall, AVG measures seen attacks, FwdTrans measures zero-day attacks, and BwdTrans measures forgetting. For all metrics, a higher positive result indicates a better performance. 

We also report the threshold-free metric Precision-Recall Area Under the Curve (PR-AUC) \cite{praucDavid06}. Since \Design{} requires selecting a threshold, PR-AUC allows us to assess model performance independently of the threshold. We choose PR-AUC over Receiver Operating Characteristic Area Under the Curve (ROC-AUC) because ROC-AUC can give misleadingly high results in the presence of class imbalance \cite{praucDavid06}.

\textbf{Hyperparameters:} 
We utilize \textit{elbow method} \cite{han2011data} for determining the number of clusters $K$. 
We set $\lambda_R$ and $\lambda_{CL}$ to 0.1, and for $m$ we use 2 after careful experimentation. For the AE modules of \Design{}, we use 4-layer MLP with 256 neurons in the hidden layers. We train it using Adam optimizer \cite{kingma2017adammethods} with a learning rate of 0.001. For PCA, we use the explained variance method and set it to 95\% \cite{rios2022incdfm}.

\textbf{Hardware:} We run our experiments on NVIDIA GeForce RTX 3090 GPU, with a AMD EPYC 7343 16-Core processor.

\subsection{Results}

\textbf{Continual Learning Comparison:} Fig.~\ref{fig:continual_methods_results} presents the results of our approach \Design{} compared with ADCN\cite{ashfahani2023unsupervised} and LwF\cite{lwf2019Li}. \Design{} shows the best performance on both seen (AVG) and unseen (FwdTrans) attacks across all datasets. \Design{} also has the highest BwdTrans on all except one dataset (UNSW-NB15). The average BwdTrans of \Design{} (0.87\%) is higher than the average BwdTrans of both ADCN (-0.06\%) and LwF (0.09\%). Notably, the BwdTrans of \Design{} is positive for three datasets. Indicating past experiences actually improve after training on future experiences for these datasets. Given the high FwdTrans as well, our approach finds features that generalize well to future experiences. 

Table~\ref{tab:improvement} shows the improvement of \Design{} over the UCL baselines on all datasets. Bold and underlined cases indicate the best and the second best improvements with respect to each metric, respectively. These improvements were calculated by comparing the performance of \Design{} to the baselines, where the improvement values represent the proportional increase over the baseline performance. We do not include BwdTrans because a proportional increase does not make sense for a metric that can be negative. \Design{} has up to $4.50\times$ and $6.1\times$ AVG improvement on ADCN and LwF, respectively. In addition, \Design{} has up to $6.47\times$ and $3.47\times$ FwdTrans improvement on ADCN and LwF. Averaged across all datasets, \Design{} shows a $1.88\times$ and $1.78\times$ improvement on AVG, and a $2.63\times$ and $1.60\times$ improvement on FwdTrans, compared to ADCN and LwF, respectively. 

Overall, these results highlight the benefit of continual ND over UCL methods for IDS. \Design{}, with its PCA-based novelty detector, excels by effectively harnessing the normal data to identify attacks. A key strength of our approach lies in the assumption that normal data forms a distinct class, while everything else is treated as anomalous. This assumption is particularly well-suited to IDS. In contrast, methods like ADCN and LwF do not make this distinction where they handle both normal and attack data similarly, limiting their ability to fully exploit the inherent structure of the data.


\begin{table}[]
\centering
\caption{\Design{} Improvement over UCL Baselines}
\label{tab:improvement}
\scalebox{1}{
\begin{tabular}{|c|c|c|c|}
\hline
Baseline      & Dataset    & AVG  & FwdTrans  \\ \hline
ADCN\cite{ashfahani2023unsupervised}  & X-IIoTID   & $\underline{2.02\times}$  & $\underline{5.00\times}$   \\ \cline{2-4} 
                                      & WUSTL-IIoT & $\mathbf{4.50\times}$  & $\mathbf{6.47\times}$   \\ \cline{2-4} 
                                      & CICIDS2017 & $1.37\times$  & $1.73\times$   \\ \cline{2-4} 
                                      & UNSW-NB15  & $1.29\times$  & $1.44\times$   \\ \hline
LwF\cite{lwf2019Li}                   & X-IIoTID   & $1.46\times$  & $1.35\times$   \\ \cline{2-4} 
                                      & WUSTL-IIoT & $\mathbf{6.11\times}$  & $\mathbf{3.47\times}$   \\ \cline{2-4} 
                                      & CICIDS2017 & $\underline{1.93\times}$  & $\underline{2.64\times}$   \\ \cline{2-4} 
                                      & UNSW-NB15  & $1.11\times$  & $1.02\times$   \\ \hline
\end{tabular}}
\end{table}

\begin{figure}
    \centering
    \includegraphics[width=0.9\linewidth]{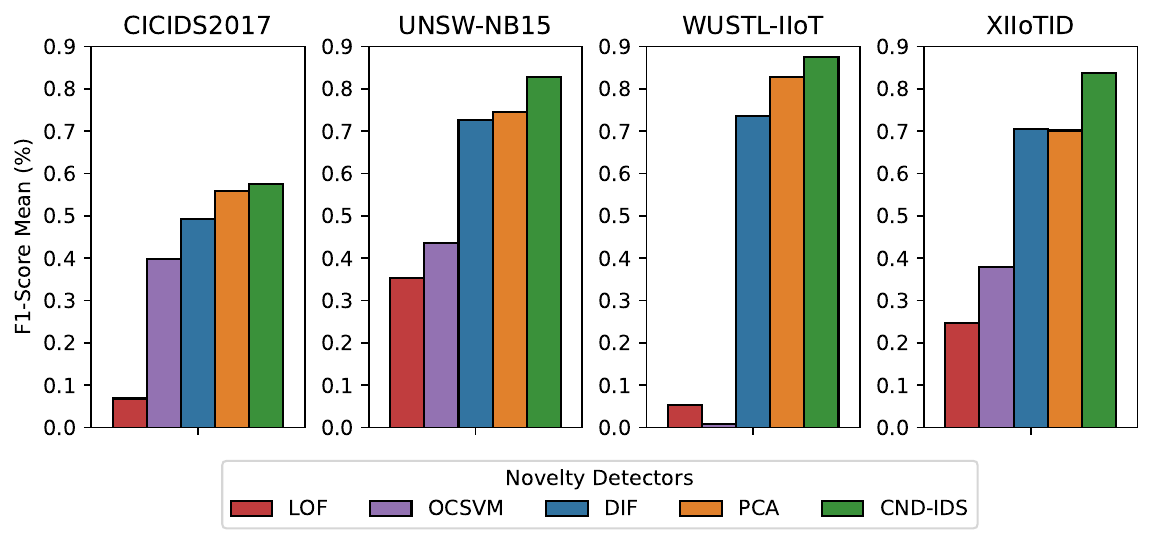}
    \caption{Average $F_1$ score on all experiences of \Design{} and novelty detection methods: LOF, OC-SVM, PCA, DIF}
    \label{fig:novelty_methods_results}
\end{figure}

\textbf{Novelty Detectors Comparison:} Fig.~\ref{fig:novelty_methods_results} compares LOF\cite{Faber_2024}, OC-SVM\cite{Faber_2024}, PCA\cite{rios2022incdfm}, and DIF \cite{xu2023deep} with \Design{} on all datasets. The average $F_{1}$ score of the novelty detection methods are compared to the AVG of \Design{}.  It can be seen \Design{} outperforms all other methods across all datasets. The two best performing methods are DIF and PCA. The average $F_{1}$ score improvement across all datasets of \Design{} is $1.16\times$ and $1.08\times$ over DIF and PCA, respectively. These results highlight the critical role of leveraging information from unsupervised data streams. Unlike these ND algorithms, \Design{} is capable of continuously learning from this unsupervised data, enabling it to enhance PCA reconstruction over time. By integrating evolving data patterns, \Design{} not only adapts to new anomalies but also improves its overall detection accuracy, demonstrating a clear advantage in dynamic environments.


\begin{figure}
    \centering
    \includegraphics[width=0.86\linewidth]{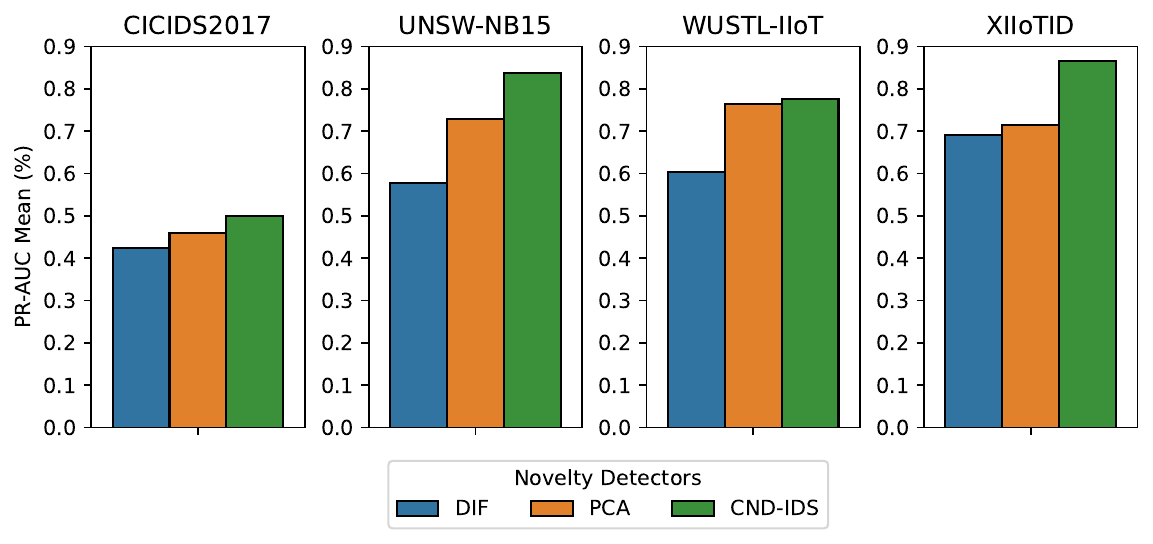}
    \caption{Thresholding Free Evaluation of \Design{}}
    \label{fig:thresholding_free}
\end{figure}


\textbf{Pre-threshold Evaluation:} While thresholding plays a crucial role in attack decision-making, evaluating model prediction performance before applying threshold is also important. The UCL algorithms (ADCN\cite{ashfahani2023unsupervised} and LwF\cite{lwf2019Li}) do not output anomaly scores because they select classes based on the closest labeled cluster. Therefore we compare against the two best ND methods: DIF\cite{xu2023deep} and PCA\cite{rios2022incdfm}. Fig.~\ref{fig:thresholding_free} presents the PR-AUC values of DIF, PCA, and \Design{}. It can be seen that \Design{} provides the best threshold free results, which aligns with the threshold-based results presented earlier. The strong performance of \Design{} in both pre-threshold and threshold-based evaluations demonstrates that the model is robust regardless of the decision threshold. 

\subsection{Ablation Study}

To demonstrate the impact of our loss function components, we perform an ablation study. Table~\ref{tab:ablation_loss} shows the results of \Design{} with each loss function removed to demonstrate their individual effectiveness. Bold and underlined cases indicate the best and the second best performances with respect to each metric, respectively. \Design{} without reconstruction loss ($L_R$) and \Design{} without cluster separation loss ($L_{CS}$) performs worse in all categories. \Design{} without both $L_R$ and continual learning loss ($L_{CL}$) actually performs better AVG but has worse BwdTrans and FwdTrans. AVG does not account for past experiences, so the significantly negative BwdTrans indicates \Design{} w/o $L_R$ and $L_{CL}$ forgets, and therefore would perform worse on those experiences in the future. This would make sense as a regularization loss to improve continual learning would slightly decrease performance in non-continual scenario. Overall \Design{} has the best results when taking every metric category into account. Notably the low BwdTrans and FwdTrans of \Design{} (w/o $L_R$) showcases how the reconstruction loss helps \Design{} generalize better to unseen and past data. This highlights the power of $L_R$ to provide good features for continual learning. 

\begin{table}[]
\caption{Ablation Study of \Design{} Loss Functions}
\label{tab:ablation_loss}
\centering
\begin{tabular}{|c|c|c|c|}
\hline
Strategy                         & AVG              & BwdTrans        & FwdTrans         \\ \hline
CND-IDS                          &\underline{76.92\%}    & \textbf{0.87\%} & \textbf{73.70\%} \\ \hline
CND-IDS (w/o $L_{CS}$)           & 66.23\%          & \underline{0.09\%}    & 70.26\%          \\ \hline
CND-IDS (w/o $L_R$)              & 72.86\%          & -5.44\%         & 67.82\%          \\ \hline
CND-IDS (w/o $L_R$ and $L_{CL}$) & \textbf{79.92\%} & -11.26\%        & \underline{71.01\%}    \\ \hline
\end{tabular}
\end{table}

\subsection{Overhead Analysis}
\begin{table}[]
\centering

\caption{Average inference time (in ms) per test sample}
\label{tab:overhead}
\scalebox{0.95}{
\begin{tabular}{|c|c|c|c|c|c|}
\hline
Strategy           & \Design{} & ADCN   & LwF    & DIF    & PCA    \\ \hline
Inference Time (ms) & \underline{0.0019}                     & 0.4061 & 0.0677 & 1.0535 & \textbf{0.0018} \\ \hline
\end{tabular}}
\end{table}
Table~\ref{tab:overhead} evaluates the inference overhead of \Design{} compared to ADCN \cite{ashfahani2023unsupervised}, LwF \cite{lwf2019Li}, DIF \cite{xu2023deep}, and PCA \cite{rios2022incdfm}. 
\Design{} offers the fastest inference time among continual learning methods. Out of novelty detection methods, \Design{} is second only to PCA. We attribute the efficiency of \Design{} to avoiding the clustering classification used by LwF and ADCN. 
The difference between \Design{} and PCA is minimal, only 0.0001 milliseconds slower, due to the additional but lightweight step of encoding the data. Considering that the average median flow duration across datasets is 27.77 milliseconds, the overhead introduced by \Design{} is negligible in the context of real-time traffic flow.

\section{Conclusion}
\label{conclusion}
IDS are essential for ensuring a robust and comprehensive security strategy. ML methods have been recently adopted for this task due to their great prediction performance. However, current ML-IDS solutions are vulnerable to under-represented attacks and constantly changing environments. Therefore, future IDS must be able to (i) generalize to unseen attacks and (ii) adapt to evolving attacks over time. To address these challenges, we present \Design{}, a continual novelty detection based IDS algorithm. \Design{} combines PCA novelty detection with a continual feature extractor. \Design{} is capable of continually learning with no attack labels, allowing it to detect attacks in a changing data stream. We have demonstrated that on realistic intrusion datasets, \Design{} is up to 6.1$\times$ more effective than SOTA unsupervised CL method for detecting seen attacks and up to 6.5$\times$ more effective at detecting future attacks. 

\section*{Acknowledgments}
This work has been funded in part by NSF, with award numbers \#1826967, \#1911095, \#2003279, \#2052809, \#2100237, \#2112167, \#2112665, and in part by PRISM and CoCoSys, centers in JUMP 2.0, an SRC program sponsored by DARPA.

\newpage
\bibliographystyle{ieeetr}
\bibliography{bibfile}

\end{document}